\documentclass[preprint]{vgtc}               % preprint

\ieeedoi{10.1109/VizSec53666.2021.00014}

\ifpdf%                                % if we use pdflatex
  \pdfoutput=1\relax                   % create PDFs from pdfLaTeX
  \usepackage{graphicx}                % allow us to embed graphics files
  \DeclareGraphicsExtensions{.pdf,.png,.jpg,.jpeg} % for pdflatex we expect .pdf, .png, or .jpg files
\else%                                 % else we use pure latex
  \usepackage{graphicx}                % allow us to embed graphics files
  \DeclareGraphicsExtensions{.eps}     % for pure latex we expect eps files
\fi%

\graphicspath{{figures/}{pictures/}{images/}{./}} % where to search for the images

\usepackage{microtype}                 % use micro-typography (slightly more compact, better to read)
\PassOptionsToPackage{warn}{textcomp}  % to address font issues with \textrightarrow
\usepackage{textcomp}                  % use better special symbols
\usepackage{mathptmx}                  % use matching math font
\usepackage{times}                     % we use Times as the main font
\usepackage{cite}                      % needed to automatically sort the references
\usepackage{tabu}                      % only used for the table example
\usepackage{booktabs}                  % only used for the table example
\usepackage[table]{xcolor}
\usepackage{graphics}
\usepackage{amssymb}

\usepackage{wrapfig}

\definecolor{c1}{rgb}{0.0, 0.3, 1.0}
\definecolor{c2}{rgb}{1.0, 0.0, 0.0}
\definecolor{c3}{rgb}{0.16, 0.5, 0.0}
\definecolor{c4}{rgb}{0.2, 0.41, 0.65}

\newcommand{\repo}{
  $\vcenter{\hbox{\includegraphics[height=1.1em]{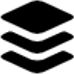}}}$
}
\newcommand{\module}{
  $\vcenter{\hbox{\includegraphics[height=1.1em]{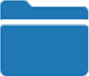}}}$
}
\newcommand{\lib}{
  $\vcenter{\hbox{\includegraphics[height=1.1em]{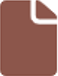}}}$
}
\newcommand{\vuln}{
  $\vcenter{\hbox{\includegraphics[height=1.1em]{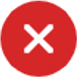}}}$
}
\newcommand{\vulncol}{
  \hspace{-0.1cm}
  $\vcenter{\hbox{\includegraphics[height=1.1em]{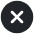}}}$
  \hspace{-0.1cm}
}
\newcommand{\link}{
  \hspace{-0.13cm}
  $\vcenter{\hbox{\includegraphics[height=1.1em]{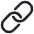}}}$
  \hspace{-0.1cm}
}
\newcommand{\graph}{
  \hspace{-0.1cm}
  $\vcenter{\hbox{\includegraphics[height=1.1em]{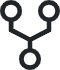}}}$
  \hspace{-0.1cm}
}

\onlineid{2128}

\vgtccategory{Research}

\vgtcinsertpkg

\title{VulnEx: Exploring Open-Source Software Vulnerabilities in Large Development Organizations to Understand Risk Exposure}

\author{Frederik L. Dennig$^1$, Eren Cakmak$^1$, Henrik Plate$^2$, and Daniel A. Keim$^1$ \\ \scriptsize $^1$University of Konstanz, Germany\thanks{e-mail: \{frederik.dennig, eren.cakmak, keim\}@uni-konstanz.de} \hfill $^2$SAP Security Research, France\thanks{e-mail: henrik.plate@sap.com}} % \textit{Member, IEEE}

\teaser{
  \centering
  \includegraphics[width=\textwidth]{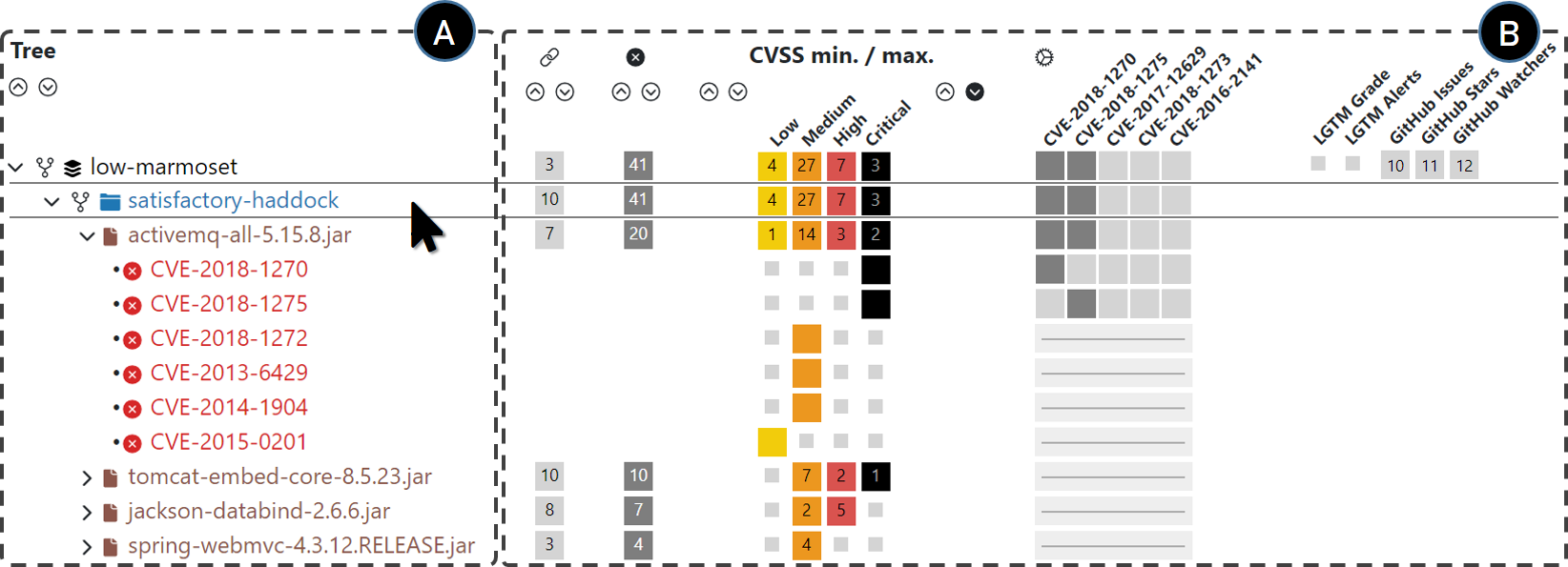}
  \caption{\textsc{VulnEx} is a tool for the investigation of exposure to open-source software vulnerabilities on an organization-wide level. The tool shows repositories, modules, libraries, vulnerabilities in a tree representation (A), and meta-information about each entry (B), such as the CVSS score. We can see that the "low-marmoset" repository is exposed to severe vulnerabilities, three critical and seven high. Two of the critical vulnerabilities are originating from the \emph{activemq-all} indicating that the library should be updated swiftly.}
  \label{fig:teaser}
}

\abstract{
The prevalent usage of open-source software (OSS) has led to an increased interest in resolving potential third-party security risks by fixing common vulnerabilities and exposures (CVEs). However, even with automated code analysis tools in place, security analysts often lack the means to obtain an overview of vulnerable OSS reuse in large software organizations. In this design study, we propose \textsc{VulnEx} (Vulnerability Explorer), a tool to audit entire software development organizations. We introduce three complementary table-based representations to identify and assess vulnerability exposures due to OSS, which we designed in collaboration with security analysts. The presented tool allows examining problematic projects and applications (repositories), third-party libraries, and vulnerabilities across a software organization. We show the applicability of our tool through a use case and preliminary expert feedback.
} % end of abstract


\begin{document}


\firstsection{Introduction}

\maketitle

%% \section{Introduction} %for journal use above \firstsection{..} instead

The extensive usage of open-source software (OSS) nowadays promotes a straightforward integration of common software features into existing applications~\cite{Pittenger2016,Sonatype2020}. However, software reuse also poses a significant risk as software with disclosed vulnerabilities is often extensively reused, affecting various applications across whole organizations~\cite{Bullough2017}. For instance, the Equifax data breach in 2017 resulted from a missed OSS package update and led to the disclosure of the private data of over 145 million U.S. citizens~\cite{Moore2017}. Hence, an organization's governance or audit system must identify the organization's overall exposure to OSS vulnerabilities.

Developers and security analysts regularly utilize automated code analysis tools to identify vulnerabilities and investigate the mitigation of OSS security risks. For example, static~\cite{Sommerville2010} and dynamic code analysis~\cite{Plate2015,Ponta2020} are applied to execute the developed code and detect inherent vulnerabilities. However, such code analysis tools heavily differ in their detection capabilities. They often only store the vulnerability metadata as text files that do not meet software developers' basic requirements, such as prioritizing the most severe vulnerabilities. Assessing the impact of software vulnerabilities is essential for organizations since the effects of exposures can vary significantly. Further, code analysis tools are usually used for single software applications and do not show the impact of OSS vulnerabilities across multiple applications in whole software organizations. Additionally, it is crucial to evaluate the quality of libraries and other dependencies if they originate from another source, such as that the source can be trusted~\cite{Sohal2018}, and that OSS developers are swift in addressing vulnerabilities~\cite{Akatsu2021}. The mentioned points are crucial for deciding whether a software development organization should use an OSS library.

We propose \textsc{VulnEx} (Vulnerability Explorer), a new tailored design to explore and assess the mitigation of OSS vulnerabilities for auditing and governance of whole software development organizations looking beyond individual applications and teams. In our user-centered design study, we designed three complementary table-based representations to identify and assess vulnerabilities across various applications. We demonstrate the applicability of our approach through use cases and initial expert feedback. \textsc{VulnEx} is open-source\footnote{\url{https://github.com/dbvis-ukon/vulnex}} and accessible online\footnote{\url{https://dennig.dbvis.de/vulnex}}. With this work, we present a first study to improve the analysis and mitigation of software vulnerabilities, especially from an organization-wide perspective. In summary, the primary contributions of this paper are: (1) A design study with problem characterization, findings, and lessons learned for the visual analysis of OSS vulnerabilities. (2) The interactive \textsc{VulnEx} analysis tool to interactively explore critical vulnerabilities. With this work, we hope to improve the analysis and mitigation of software vulnerabilities by addressing the need for an analysis tool for auditing entire software development organizations.

\section{Related Work}
Software visualizations provide a comprehensive overview of complex systems, such as program structures, execution behavior, and the development process~\cite{Diehl2007}. These visualizations are also useful to investigate security aspects, e.g., SecSTAR~\cite{Fang2012} automatically generates execution diagrams to examine, debug, and test software applications. For an overview of software visualization research, refer to the reviews of Wagner et al.~\cite{Wagner2015}, and Chotisarn et al.~\cite{Chotisarn2020}.

In software security visualization, some approaches for vulnerability exploration have been proposed. Harrison et al.~\cite{Harrison2012} proposed the Nessus vulnerability visualization (NV) to discover and analyze network vulnerabilities of Nessus scans. The system simplifies and displays vulnerability assessment results to support security analysts, using zoomable treemap visualization with linked histograms. In a similar context, Angelini et al.~\cite{Angelini2019Valnus} proposed Vulnus, which aims to increase situational awareness of security managers by visually analyzing vulnerability spreads in computer networks.  Furthermore, CVExplorer~\cite{Pham2018} is a visual analytics system for analyzing vulnerability reports and enhancing network security using three linked views. These vulnerability systems differ from our approach since they primarily focus on exposing computer network vulnerabilities. Moreover, Goodall et al.~\cite{Goodall2010} proposed a system to explore vulnerabilities and code weaknesses in software development. The goal is to help users understand their code's security status by displaying code vulnerabilities using an aggregated block metaphor for each file. Goodall et al.~\cite{Goodall2010} approach focuses on identifying false positives, which we reduce in our application by checking whether third-party vulnerabilities are reachable. Assal et al.\cite{Assal2016} presented Cesar, a collaborative code analysis system to reduce vulnerabilities and improve code security. The authors utilize a treemap visualization to help security experts and developers collaboratively explore static-code analysis methods' results. The treemap visualization displays a software package, and each leaf node shows a class file. Angelini et al.~\cite{Angelini2019SymNav} presented a visual analytics approach to assist users in exploring program execution, describing in a use case of the detection of single vulnerabilities. However, the system is mainly targeted to investigate symbolic execution engine data. Recently, Alperin et al.~\cite{Alperin2020} presented a study for the interpretable visual assessment of vulnerabilities. In their study, the authors focus on local explanations for predictive vulnerability analysis.

In summary, the listed approaches focus on exploring network vulnerabilities and improving the code security of individual software packages, such as investigating potential false positives. In contrast, we propose an initial approach that provides an overview of entire software development organizations. Our design study focuses mainly on the visual analysis of OSS vulnerabilities by supporting auditing teams in assessing OSS dependencies through table-based views to evaluate vulnerabilities in large software organizations.

\section{Problem Statement}

The main goal of this work is to design visualizations to explore security risks in large software organizations. We gathered knowledge about the domain and user requirements in three interviews with two security analysts and a software developer from SAP. The interviews provided valuable insight into the daily workflows and challenges faced by security analysts regarding vulnerability assessment.

\smallskip

\noindent
\textbf{Application Background:}
The essential user task is to understand the overall risk exposure of large development organizations, e.g., commercial software vendors or open-source foundations, due to the consumption of open-source components in a considerable number of development projects or applications. During software development, projects are regularly scanned with code analysis tools.  At SAP, the developers regularly utilize \emph{Eclipse Steady}\footnote{\url{https://github.com/eclipse/steady}}~\cite{Ponta2020}, which supports static and dynamic analysis to detect and assess vulnerabilities. \emph{Eclipse Steady} scans projects for Common Vulnerabilities and Exposures (CVE), which have a unique identifier in the National Vulnerability Database. \emph{Eclipse Steady} displays the Common Vulnerability Scoring System (CVSS) score to indicate the severity of identified security vulnerabilities. However, the CVSS score only captures the vulnerability severity. Organizations require complementary information from other sources to evaluate the general software quality of the most-used libraries and determine whether they have sufficient quality. The identification of low-quality libraries is a prerequisite for follow-up decisions. However, the visual exploration of applications, consumed libraries, and related vulnerabilities on an organizational level are not supported by any of the tools available to date.

\smallskip

\noindent
\textbf{Requirements:} From the interviews and further discussions with domain experts, we derived the following requirements for our tool aimed at the organization-wide analysis of software vulnerabilities.

\noindent
\emph{(R1) Repositories} -
The tool should provide views to detect vulnerable repositories and projects to apply countermeasures, such as training weaker teams and reallocating resources. For this, repositories need to be represented in a comparable way to estimate relevance and understand how they compare against each other.

\noindent
\emph{(R2) Libraries} -
Software projects potentially depend on vulnerable libraries, which have to be updated. Thus, the tool needs to convey the overall exposure and allow for the inspection of specific bugs.

\noindent
\emph{(R3) Vulnerabilities} -
Vulnerabilities need to be explored to address specific exploited known vulnerabilities, e.g., OSS vulnerabilities prominently discussed in mainstream media, where organizations may be required or expected to make a statement whether and which of their applications are affected. Thus, the tool needs to enable users to find specific bugs with a high-security risk.

\noindent
\emph{(R4) Vulnerability Severity} -
Vulnerabilities can have different effects depending on the severity and how many projects the origin is, and thus need to be prioritized accordingly. Therefore, the tool needs to show the impact of specific bugs on the organization's codebase.

\section{Table-based Vulnerability Exploration}

In a two-year process, we applied the guidelines by Chen et al.~\cite{Chen2019} to perform our design study. Our tool covers the pipeline from scanning to repairing or mitigating a vulnerability. The overall workflow of \textsc{VulnEx} follows the KDD pipeline~\cite{Fayyad1996}. The workflow of \textsc{VulnEx}: (1) The security analyst starts a scan of the source code of all software projects. (2) He then selects a type of analysis target, i.e., repositories, libraries, or vulnerabilities, choosing between overviews. (3) Then, the analyst defines criteria he is interested in, i.e., the number or severity of a bug allowing for filtering. (4) The analyst observes the findings and determines their relevance by drilling down to the specific issue. (5) In case of a relevant finding, the analyst can start a repair or mitigation process; this is supported by the detailed report of \emph{Eclipse Steady}. Thus, we follow Shneidermans' mantra: \emph{Overview first, zoom and filter, then details-on-demand}~\cite{Shneiderman1996}. In \autoref{fig:teaser} we show the dependency tree (A), allowing users to explore the hierarchy of the software project and the vulnerability information (B) to give insight into the exposure to vulnerabilities.

\subsection{Dependency Tree}

The dependency tree representation in \autoref{fig:teaser} (A) shows the relationship of all repositories~\hspace{-0.05cm}\repo\hspace{-0.05cm}, modules \hspace{-0.05cm}\module\hspace{-0.05cm}, libraries \hspace{-0.05cm}\lib\hspace{-0.05cm}, and bugs\hspace{-0.05cm}\vuln\hspace{-0.05cm}. \textsc{VulnEx} is inspired by the tree+table approach by Nobre et al.~\cite{Nobre2019Lineage,Nobre2019Juniper}. We choose this because tree structures are common and known by domain experts and allow us to leverage the hierarchy inherent in software projects while supplying additional information about vulnerabilities, keeping a high level of detail. The tree representation allows for the analysis of vulnerabilities in three ways.

\noindent
\textbf{Repository-centered:} \repo $\rightarrow$ \module $\rightarrow$ \lib $\rightarrow$ \vuln \\
This order of levels allows for a repository-focused analysis. Starting with a repository, then showing information about modules and sub-modules, enabling analysts to locate severe vulnerabilities. If a module uses a vulnerable library, this can be quickly detected. Finally, we show the vulnerabilities caused by a library, allowing for detailed analysis and estimation of the impact.

\noindent
\textbf{Library-centered:} \lib $\rightarrow$ \vuln $\rightarrow$ \repo $\rightarrow$ \module \\
Beginning with a library, displaying its vulnerabilities allows analysts to estimate the risk associated with a library. If a repository uses a vulnerable library, this repository is shown on the next level. Finally, we present the associated module or sub-module exposed to the CVE of that library, allowing for inspecting it in detail.

\noindent
\textbf{Bug-centered:} \vuln $\rightarrow$ \lib $\rightarrow$ \repo $\rightarrow$ \module

\noindent
Starting with a CVE, then showing the affected library allows analysts to find specific bugs quickly. If a repository uses a vulnerable library, this repository is shown on the next level. Finally, we display the associated module or sub-module impacted by the CVE of that library, allowing for detailed analysis.

\subsection{Vulnerability Information}

We provide additional information about the vulnerabilities of a repository, module, and libraries, shown in \autoref{fig:teaser} (B), through which we support the detection and analysis of critical vulnerabilities, as well as the assessment of the quality of OSS dependencies, e.g., LGTM grade and score. The \link column shows the number of entities on the next level of the tree, indicating the number of related entities on the following hierarchy level. The \vulncol column shows the number of vulnerabilities a repository, module, or library is exposed to. The absence of an element indicates that the information is not available or applicable to the entity of the row.

\smallskip

\begin{wrapfigure}[5]{r}{1.3cm} % lines, orientation, size
    \vspace{-10pt} % translation for margins
    \hspace{-7mm}
    \includegraphics[trim=0 105 0 0,clip,width=1.9cm]{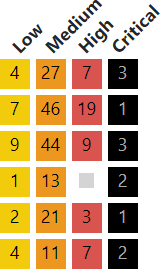}
\end{wrapfigure}
\noindent
\textbf{CVSS Score:} The CVSS score column shows the number of CVEs with a given score. We use the common classification: Low (0.1 - 3.9), Medium (4.0 - 6.9), High (7.0 - 8.9), Critical (9.0 - 10.0), which is also mirrored in the coloring from the \emph{National Security Database}\footnote{\url{https://nvd.nist.gov/}}. The number in each square encodes the number of occurrences in the given range.

\begin{wrapfigure}[2]{r}{2.6cm} % lines, orientation, size
    \vspace{-12pt} % translation for margins
    \hspace{-6.5mm}
    \includegraphics[width=3.0cm]{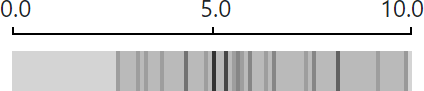}
\end{wrapfigure}
\noindent
To inspect the distribution of CVSS scores in a more finely-grained way, we offer a representing of each CVE and its CVSS score with its precise value. It also indicates the range of CVSS scores.

\smallskip

\begin{wrapfigure}[6]{r}{1.3cm} % lines, orientation, size
    \vspace{-12pt} % translation for margins
    \hspace{-7mm}
    \includegraphics[trim=0 25 0 0,clip,width=1.9cm]{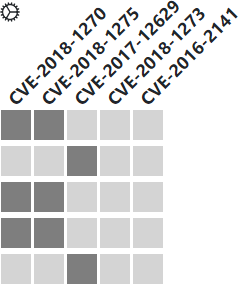}
\end{wrapfigure}
\noindent
\textbf{CVE Matrix:} The CVE matrix indicates the presence of a specific vulnerability. Each column shows the presence of a CVE with dark gray squares, while a light gray square indicates the absence of the CVE. We adopted this encoding from Nobre et al.~\cite{Nobre2019Juniper}. Columns can be added and removed to highlight specific CVEs dependent on the user. The CVE matrix allows users to get an overview of the presence of specific vulnerabilities in repositories, modules, and libraries. It also enables the analysis of the co-occurrence of CVEs.

\smallskip

\begin{wrapfigure}[5]{r}{2cm} % lines, orientation, size
    \vspace{-12pt} % translation for margins
    \hspace{-7mm}
    \includegraphics[trim=0 105 0 0,clip,width=2.4cm]{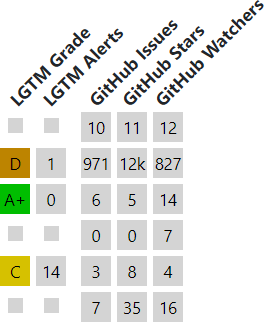}
\end{wrapfigure}
\noindent
\textbf{Meta-Information:} We provide additional information from LGTM\footnote{\url{https://lgtm.com/}}, a code analysis platform, and GitHub\footnote{\url{https://github.com/}}. The columns describe the LGTM grade, LGTM score, GitHub issues, GitHub stars, GitHub watchers. The LGTM grade and score provide an additional measure for the quality of software artifacts. The number of GitHub issues provides an indicator for active development, while the GitHub stars and watchers provide an indicator for the popularity of a repository.

\smallskip

\begin{wrapfigure}[5]{r}{2.6cm} % lines, orientation, size
    \vspace{-12pt} % translation for margins
    \hspace{-8mm}
    \includegraphics[width=3.2cm]{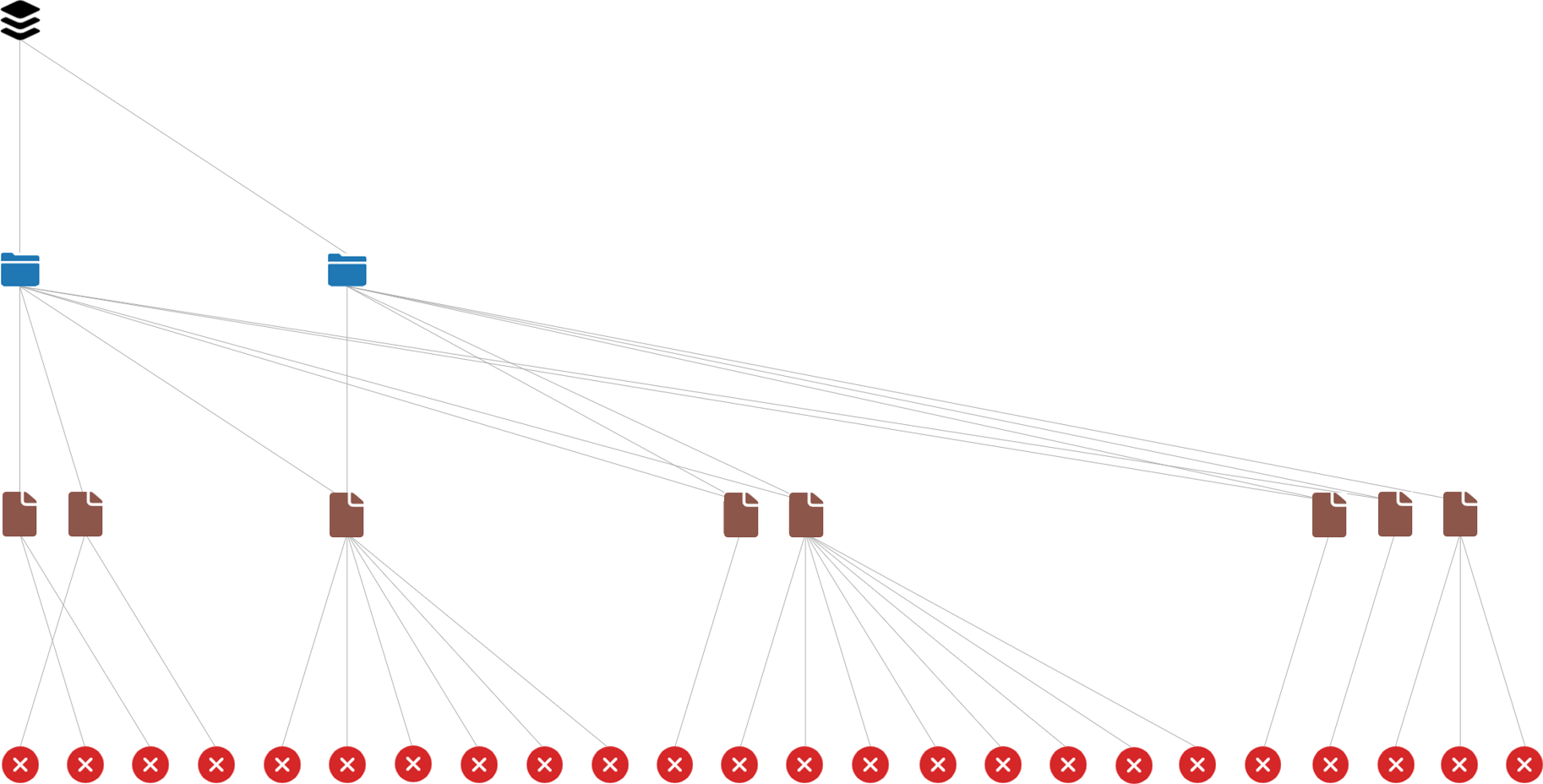}
\end{wrapfigure}
\noindent
\textbf{Dependency Graph:} The user can view the structure of a software project by clicking \hspace{-0.05cm}\graph\hspace{-0.05cm}. The repository is shown at the top, its modules and libraries in the middle, and the bugs at the bottom.

\smallskip

\subsection{Filter and View Options}
Based on expert feedback, we offer filter options to reduce the number of entries in the table and allow for a focused analysis. The user can search for a name of a given repository, library, or bug. We enable users to filter by the minimum and the maximum number of dependencies, vulnerabilities, and the CVSS score. Users can hide all repositories and modules that do not contain any vulnerability, as well as CVEs without a CVSS score.

\section{Evaluation}

\subsection{Use case: Eclipse GitHub Repositories}

We analyze all public GitHub repositories of the \emph{Eclipse Foundation}\footnote{\url{https://github.com/eclipse}} that are \emph{Apache Maven}\footnote{\url{https://maven.apache.org/}} projects in the \emph{Java}\footnote{\url{https://docs.oracle.com/en/java/index.html}} programming language. All projects are scanned using the \emph{Eclipse Steady} tool. We scanned these repositories from January 21 to February 4, 2020. This yields a total of 295 projects that we analyze for common libraries and vulnerabilities. We replace the original repository and module names with pseudonyms not to blame the individual projects. At SAP, the analysis of an individual application follows a defined process, starting from the automated scanning with tools like \emph{Eclipse Steady} to discover vulnerable open-source dependencies, the assessment of findings by a security expert, and finally, depending on the assessment result, the remediation of the vulnerability or the dismissal of the finding. However, open-source software analysis across multiple applications for an entire organization does not follow a defined process. To show the usefulness of \textsc{VulnEx} we analyze the gathered data to answer the following four questions. Questions (Q1-Q3) are examples for exploratory analysis, while (Q4) addresses a need when vulnerabilities in open-source components get a lot of public attention, even in mainstream media. In such cases, commercial vendors like SAP are expected to state to what extent and which of their applications are affected by a given vulnerability. Thus, we include (Q4) as a search task.

\smallskip

\noindent
\emph{(Q1) Which repositories contain most severe vulnerabilities?}

\noindent
Security analysts utilize the \emph{Repository Table} to analyze all repositories, depicted in~\autoref{fig:teaser}. They find the "low-marmoset" repository, which has three critical bugs. We can see that all critical vulnerabilities are in the "satisfactory-haddock" module by expanding the entry. They inspect the module and see that the \emph{tomcat-embed-core} library contains \emph{CVE-2018-8014} and \emph{activemq-all} contains \emph{CVE-2018-1270} and \emph{CVE-2018-1270}. They find that all three CVEs are critical, which should be addressed promptly.

\smallskip

\noindent
\emph{(Q2) Which dependencies contain severe vulnerabilities and are often used across different applications?}

\begin{figure}[t] %bh
    \centering
    \includegraphics[width=0.9\linewidth]{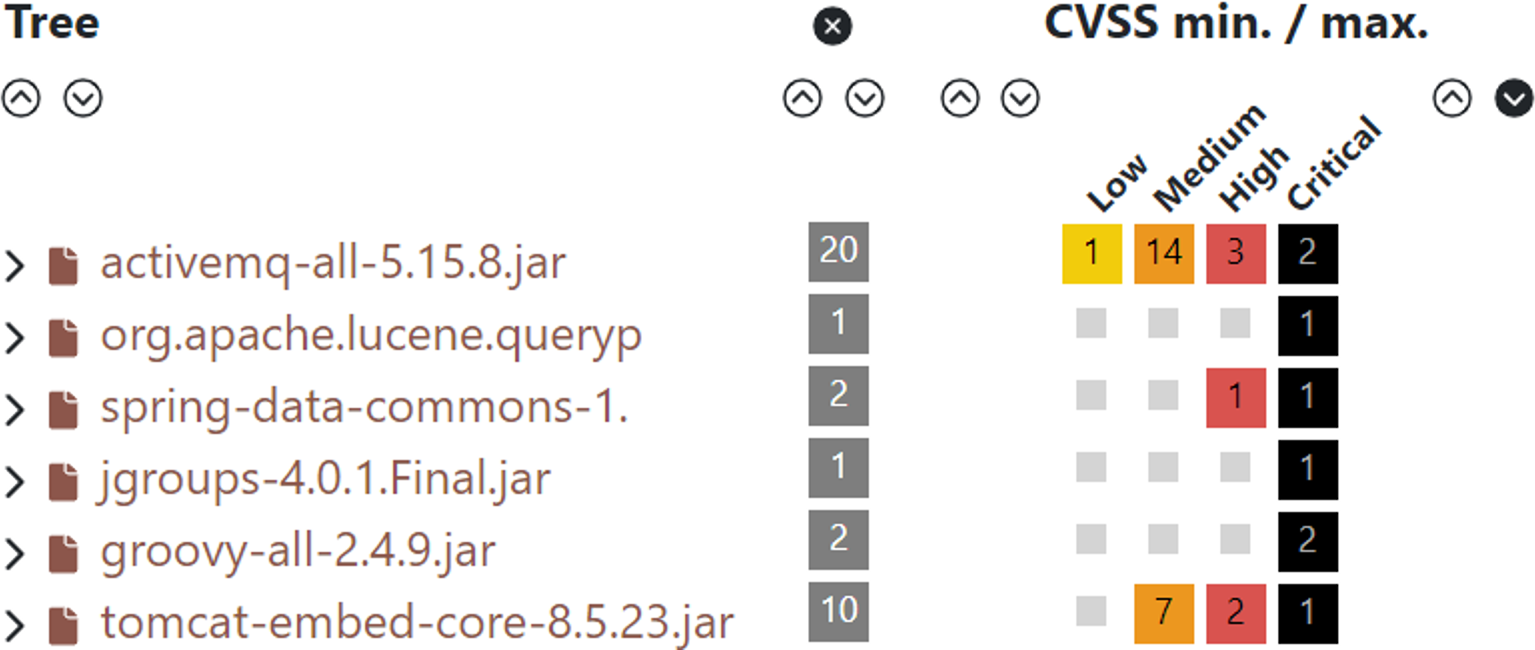}
    \caption{The analyst detects the most vulnerable libraries. 
    \emph{activemq-all} contains one low, 14 medium, three high, and two critical severity vulnerabilities, affecting 20 repositories.}
    \label{fig:q3}
\end{figure}

\noindent
The security analysts use the \emph{Library Table} (see~\autoref{fig:q3}). They sort the table by the most severe vulnerability. The libraries \emph{activemq-all}, \emph{org.apache.lucene.queryparser}, \emph{spring-data-commons}, \emph{jgroups}, \emph{groovy-all}, and \emph{tomcat-embed-core} all contain critical bugs.

\smallskip

\noindent
\emph{(Q3) Which severe vulnerabilities are present?}

\begin{figure}[t] %bh
    \centering
    \includegraphics[width=0.75\linewidth]{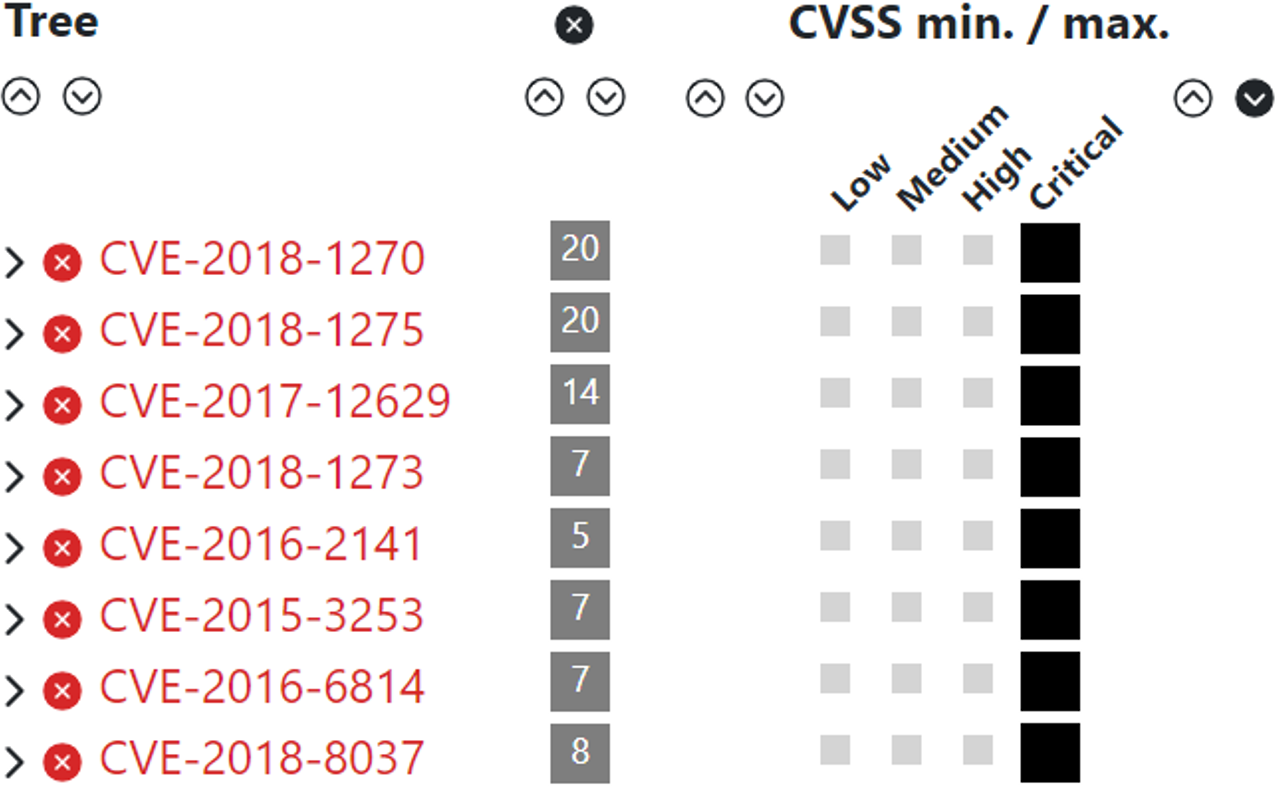}
    \caption{The analyst found eight critical vulnerabilities. \emph{CVE-2018-1270} affects 20 repositories and has a critical severity.}
    \label{fig:q4}
\end{figure}

\noindent
Using the \emph{Bug Table}, the analysts find that eight critical bugs (see~\autoref{fig:q4}) are present, one in \emph{activemq-all} affecting 20 repositories, one in \emph{org.apache.lucene.queryparser} affecting 14 repositories, one in \emph{spring-data-commons} affecting seven repositories, one in \emph{jgroups} affecting five repositories, two in \emph{groovy-all} affecting seven repositories, and one in \emph{tomcat-embed-core} affecting eight repositories. They remark that these six libraries should be updated and fixed or replaced swiftly since they contain critical vulnerabilities.

\smallskip

\noindent
\emph{(Q4) Are specific vulnerabilities present in any of the projects?}

\noindent
Analysts searched for the oldest bug for the severities medium, high, and critical. For this task, they use the \emph{Bug Table}. CVEs encode the years that they were detected. To find the oldest unfixed bug, they searched for the different years before 2019. They found \emph{CVE-2009-2625}, a medium severity bug,  present in \emph{org.apache.xerces}, which affects 27 repositories. The oldest high severity bug they found was \emph{CVE-2013-1768} in \emph{openjpa-asm-shaded}, affecting three repositories and \emph{CVE-2015-3253}, a critical bug, affecting seven repositories.

\subsection{Preliminary User Feedback}

We conducted an initial preliminary user feedback session with three software security analysts from SAP. All three participants (P1–P3) have five to ten years of experience in software security and work in dedicated security teams. Two participants support developers of mature applications regarding software security, including assessing the relevance and severity of vulnerabilities in open-source components. One participant acted as program manager for open-source security and DevSecOps, determining requirements, developing tools, and standardizing the secure consumption and publication of open-source components at SAP. We adopted the pair analytics guidelines of Kaastra and Fisher~\cite{kaastra2014field} to structure our interviews conceptually. During the one-hour interviews, we gathered regular user tasks, related employed visual interfaces, familiarity with information visualization for cyber-security, and afterward reviewed and compared in a live session their initial expectations to the proposed \textsc{VulnEx} tool.

All three participants approved the usefulness of \textsc{VulnEx} to visually explore the use of open-source libraries in large software organizations. P1 and P2 appreciated that the tool displays how often libraries and their potential vulnerabilities are used in the whole organization. P3 liked that the CVE matrix displays the top five bugs in the organization as it highlights the affected packages, including other prevalent vulnerabilities with their CVSS scores. Overall, all participants believed that \textsc{VulnEx} tool helps explore software organizations' vulnerabilities from different perspectives, such as in repository, library, and bug table views.  

The participants expressed some concerns and outlined some shortcomings of our tool. P1 suggested adding additional information about the open-source libraries to the tool, such as short descriptions of the main functionality and purpose of the library. The participant argued that keeping track of each library's functionality without such additional information remains challenging due to the sheer number of 3rd party libraries. P2 emphasized that the current visual representations might not scale to large-scale organizations, e.g., organizations with more than 1000 repositories. P2 also proposed to enable the annotation of individual repositories, libraries, and bugs. Such annotations let analysts search for particular vulnerabilities and guide the auditing team to potential known solutions. P3 emphasized that his focus is heavily on vulnerabilities with critical CVSS scores above $9.0$ that need to be resolved within several hours. Therefore, P3 suggested focusing on such vulnerabilities and recommending appropriate counter-measures. All participants suggested including potential solutions to resolve the vulnerabilities.

\section{Discussion and Conclusion}

We found that three security experts approved the usefulness of \textsc{VulnEx}. The experts found the different task-focused views useful. We learned that more detailed representations were less preferred. The domain experts had an easier time working with the categories low, medium, high and critical, rather than the precise values of the heatmap visualization. The CVE matrix gives a helpful overview of specific vulnerabilities. All vulnerability analysis tools at SAP focus on individual applications. Thus, we present \textsc{VulnEx} supporting organization- and enterprise-wide decision making. In terms of scalability, we performed our analyses on all public GitHub repositories of \emph{Eclipse Foundation}. Therefore, we argue that \textsc{VulnEx} can be used for large software organizations since few organizations have more projects than the \emph{Eclipse Foundation}.

We plan to address the feedback from the security experts by including a method to annotate repositories, modules, libraries, vulnerabilities and provide additional information for each item which could be taken from \emph{libraries.io} or comparable online services. We also plan to include the temporal component, analyzing multiple "snapshots" to compare projects and understand how the organization's risk exposure develops over time. Another goal is to extend \textsc{VulnEx} for the assessment of libraries before choosing a specific one and provide a feedback loop to inform the open-source community and add the vulnerability information to the original repository. We also plan to evaluate \textsc{VulnEx} with experts external to the design process. Our approach is transferable to other organizations and open-source vulnerability analysis tools, but \textsc{VulnEx} is currently limited to the import and processing of scan results from \emph{Eclipse Steady} allowing for the analysis of \emph{Java} and \emph{Python} code.

Determining the impact of vulnerabilities on software organizations is challenging due to the missing aggregation of software analysis results. As a solution, we propose the \textsc{VulnEx} (Vulnerability Explorer) tool, which we designed in a user-focused design process, which allows analysts to detect severe and relevant vulnerabilities and determine impacted libraries, modules, and repositories.

\acknowledgments{We thank the anonymous reviewers for their valuable feedback. This project has received funding from the European Union's Horizon 2020 research and innovation programme under grant agreement No 830892.}

\bibliographystyle{abbrv-doi-hyperref-narrow}

\bibliography{0-references}
\end{document}